\documentclass[twocolumn,eqsecnum,floats,aps,nofootinbib,showpacs]{revtex4}
\usepackage{latexsym}
\usepackage{amssymb}

\renewcommand{\l}{\left(}
\renewcommand{\r}{\right)}
\usepackage{amsmath, amsthm, amssymb}

\def\be{\begin{equation}}
\def\ee{\end{equation}}
\def\beq{\begin{equation*}}
\def\eeq{\end{equation*}}
\def\ba{\begin{aligned}}
\def\ea{\end{aligned}}

\def\ov{\overline}

\begin{document}
\title{Modified coupling procedure for the Poincar{\'e} gauge theory of gravity}

\author {Marcin Ka\'zmierczak}
\email{marcin.kazmierczak@fuw.edu.pl}
\affiliation{Institute of Theoretical Physics, Uniwersytet Warszawski, Ho\.{z}a 69, 00-681 Warszawa, Poland}
\begin{abstract}
The minimal coupling procedure, which is employed
in standard Yang--Mills theories, appears to be ambiguous in the case
of gravity. We propose a slight modification of this procedure, which removes the ambiguity. Our modification justifies
some earlier results concerning the consequences of the Poincar{\'e}
gauge theory of gravity. In particular, the predictions of the Einstein--Cartan
theory with fermionic matter are rendered unique. 
\end{abstract}
\pacs{04.50.Kd, 04.40.-b, 11.30.Er, 11.15.-q}
\maketitle
\section{Introduction}\label{section1}
Since the introduction by Yang and Mills of the non--Abelian gauge
theories \cite{YaMi}, attempts have been undertaken of describing all the known
interactions as emerging from the localization of some fundamental
symmetries of the laws of physics. It is now clear that all the
non--gravitational fundamental interactions can be successfully given
such an interpretation. The Yang--Mills (YM) theories constitute a formal basis
for the standard model of particle physics. Although the attempts to
describe gravity as a gauge theory were initiated by Utiyama \cite{Ut}
within a mere two years after the pioneering work of Yang and Mills,
the construction of this theory seems yet not to be satisfactorily
completed. 

If a field theory in Minkowski space is given, this theory being symmetric
under the global action of a representation of a Lie group, the
natural way to introduce the corresponding interaction within the
spirit of YM is to apply the minimal coupling
procedure (MCP). However, trying to
apply MCP in order to pass from a field theory in flat space to
a Riemann--Cartan (RC) space (i.e. a manifold equipped with a metric tensor and a
metric connection) results in difficulties. This is because
adding a divergence to the flat space Lagrangian density, which is a
symmetry transformation, leads to the non--equivalent
theory in curved space after MCP is applied. Although this problem was
observed already by Kibble\cite{Kib1}, it has been largely ignored in the
subsequent investigations concerning EC theory. The resulting
ambiguity can be physically important for the standard
Einstein--Cartan theory and its modifications \cite{Kazm1,Kazm2}. It seems that MCP should
be somehow modified for the sake of connections with torsion, so that
it gives equivalent results for equivalent flat space
Lagrangians. An attempt to establish such a modification was made by Saa \cite{Saa1,Saa2}. Unfortunately, Saa's
solution results in significant departures from general relativity, which
seem incompatible with observable data \cite{BFY}\cite{FY},
unless some additional assumptions of rather artificial nature are made, such as demanding
a priori that part of the torsion tensor vanish \cite{RMAS}. The main
purpose of this paper is to introduce an alternative modification of
MCP, which also eliminates the ambiguity. Unlike Saa's proposal, our
approach does not lead to radical changes in the predictions of the
theory. In the case of gravity with fermions, the procedure simply
justifies the earlier results of \cite{HD,HH,Ker,Rumpf,PR}. These results were
obtained partly `by chance', as the flat space Dirac Lagrangian was
randomly selected from the infinity of equally good possibilities. 
\section{The gauge approach to gravity and the ambiguity of minimal coupling}\label{section2}
Let us recall the classical formalism of a YM theory of
a Lie group $G$. Let
\be\label{S}
S[\phi]=\int\mathcal{L}\l{\phi,\partial_{\mu}\phi}\r
d^4x=\int\mathfrak{L}\l{\phi,d\phi}\r
\ee
represent the action of a field theory in Minkowski space $M$. Here
$\mathcal{L}$ is a Lagrangian density and $\mathfrak{L}$ a Lagrangian four--form. Assume
that $\mathcal{V}$ is a (finite dimensional) linear space in which fields $\phi$ take their values,
$\phi:M\rightarrow \mathcal{V}$, and $\pi$ is a representation of $Lie(G)$ on $\mathcal{V}$.
Let $\rho$ denote the corresponding representation of the
group\footnote{More precisely, in a generic case $\rho$ is a
  representation of the universal covering group of $G$, which may not
  be a representation of $G$ itself.},
$\rho\l{\exp (\mathfrak{g})}\r=\exp\l{\pi(\mathfrak{g})}\r$. 
If the Lagrangian four--form is invariant under its global action $\phi\rightarrow\phi'=\rho(g)\phi$,
one can introduce an interaction associated to the symmetry group
$G$ by allowing the group element $g$ to depend on space--time point
and demanding the theory to be invariant under the local action of
$G$. This can be most easily achieved by performing the replacement
\be\label{MCPYM}
d\phi\rightarrow \mathcal{D}\phi=d\phi+\mathcal{A}\phi ,
\ee
where $\mathcal{A}$ is a $Lin(\mathcal{V})$--valued one--form field on
$M$ ($Lin(\mathcal{V})$ being the set of linear maps of  $\mathcal{V}$
into itself) which transforms under the local action of $G$ as
\be\label{gauge}
\mathcal{A}\rightarrow\mathcal{A}'=\rho(g)\mathcal{A}\rho^{-1}(g)-d\rho(g)\rho^{-1}(g) .
\ee
In the standard YM one requires that
$\mathcal{A}$ takes values in a linear subspace
$Ran(\pi):=\{\pi(\mathfrak{g}):\mathfrak{g}\in Lie(G)\}\subset
Lin(\mathcal{V})$, but this requirement is not necessary to make the
action invariant under local transformations. We shall adopt a more general approach, in which
$\mathcal{A}$ assumes the form
\be\label{AandB}
\mathcal{A}=\mathbb{A}+\mathbb{B}(\mathbb{A},e) ,
\ee
where $\mathbb{A}$ is the usual YM connection taking values
in $Ran(\pi)$ and transforming according to (\ref{gauge}), $e$ denotes
an orthonormal basis of one--form fields serving physically as a
reference frame at each point of space--time\footnote{In the case of
  non--gravitational interactions, this frame can be fixed once and for
all and the dependence on $e$ does not have to be considered. In the
case of gravity, an orthonormal cotetrad can be constructed from
the Poincar{\'e} gauge fields. It could be then interpreted as a part of
$\mathbb{A}$, if the representation $\pi$ of the Poincar{\'e} algebra
was faithful. However, physical matter fields usually transform
trivially with respect to translations and representations $\pi$ are not
faithful. it is therefore necessary to assume separately that $\mathbb{B}$
depends on $e$.},
$\mathbb{B}(\mathbb{A},e)$ is a $Ran(\pi)^{\perp}$--valued one--form on $M$. Here
$\perp$ denotes the orthogonal complement with respect to some natural
scalar product on $Lin(\mathcal{V})$. The simplest candidate for this
scalar product is $\langle\langle X,Y
\rangle\rangle=trace\l{X^{\dag}Y}\r$, where $\dag$ stands for Hermitian
conjugation of a matrix. However, if $\mathcal{V}$ admits a $\rho$--invariant scalar product
$\langle,\rangle_{\rho}$, such that $\forall v,w\in\mathcal{V}$, $g\in G$, 
$\langle\rho(g)v,\rho(g)w\rangle_{\rho}=\langle v,w\rangle_{\rho}$,
then the use of the induced scalar product
$\langle\langle,\rangle\rangle_{\rho}$ on $Lin(\mathcal{V})$
satisfying $\langle\langle\rho(g)X\rho^{-1}(g),\rho(g)Y\rho^{-1}(g)\rangle\rangle_{\rho}=\langle\langle
X,Y\rangle\rangle_{\rho}$ may seem esthetically more appealing. This product may not be positive--definite,
but if the subspace $Ran(\pi)\subset Lin(\mathcal{V})$ is
nondegenerate with respect to $\langle\langle,\rangle\rangle_{\rho}$,
then the space of linear maps decouples into a simple sum
$Lin(\mathcal{V})=Ran(\pi)\oplus Ran(\pi)^{\perp}$ and hence
$\mathbb{A}$ and $\mathbb{B}(\mathbb{A},e)$ are uniquely determined by
$\mathcal{A}$.

In order not to introduce additional fields,
$\mathbb{B}$ is required to be determined by $\mathbb{A}$ and $e$. In order not to destroy the transformation
law (\ref{gauge}), it is also required that
$\mathbb{B}(\mathbb{A}',e')=\rho(g)\mathbb{B}(\mathbb{A},e)\rho^{-1}(g)$.
Our final requirement is that the coupling procedure thus obtained by
free of the ambiguity corresponding to the possibility of the addition
of a divergence to the initial matter action. It is
remarkable that in the case of the gravitational interaction and fermions these ideas, together with the natural requirement that the Leibniz rule holds for vector fields composed of spinors, fix the
form of $\mathbb{B}(\mathbb{A},e)$ (up to terms that can be absorbed by other known fundamental interactions and do not influence the resulting connection on the base manifold), as we will see below. All
the constructions of YM can be accomplished in terms of $\mathbb{A}$
and its curvature
$\mathbb{F}=d\mathbb{A}+\mathbb{A}\wedge\mathbb{A}$. The role of
$\mathbb{B}$ is only to modify the coupling procedure such that it is
unique.

In the case of gravity, it is not sufficient to perform the
replacement (\ref{MCPYM}) -- one needs also to replace the Minkowski
space (holonomic) basis of orthonormal one--forms $dx^{\mu}$ by the
cotetrad $e^a$ and redefine the geometric structure of the base
manifold such that the original Minkowski space $M$ becomes the RC space $\mathcal{M}(e,\omega)$ (here $\omega$ is a spin--connection that can be extracted out of $\mathbb{A}$). We shall use the
Dirac field case as an instructive example. In particle physics, the most
frequently used Lagrangian four--form for the Dirac field is
\be\label{LF0}
\ba
&\mathfrak{L}_{F0}=
-i\l{\star dx_{\mu}}\r \wedge \ov{\psi}\gamma^{\mu}
  d\psi-m\ov{\psi}\psi\, d^4x \\
&=\ov{\psi}\l{i\gamma^{\mu}\partial_{\mu}-m}\r\psi\, d^4x . 
\ea
\ee
Here $\gamma^{\mu}$ are the Dirac matrices obeying 
$\gamma^{\mu}\gamma^{\nu}+\gamma^{\nu}\gamma^{\mu}=2\eta^{\mu\nu}$,
where $\eta=diag(1,-1,-1,-1)$ is the Minkowski matrix, and
$\ov{\psi}:={\psi}^{\dagger}\gamma^0$, where ${\psi}^{\dagger}$ is
a Hermitian conjugation of a column matrix (think of $\psi$ as a
column of four complex--valued functions on space--time).
This four--form is invariant under the global action of the Poincar{\'e} group
\beq
\ba
&x^{\mu}\rightarrow x'^{\mu}={\Lambda^{\mu}}_{\nu} x^{\nu}+a^{\mu}, \quad \psi\rightarrow {\psi}'=S(\Lambda)\psi, \\
&S(\Lambda(\varepsilon)):=\exp\l{-\frac{i}{4}\varepsilon_{\mu\nu}\Sigma^{\mu\nu}}\r,
 \ \Sigma^{\mu\nu}:=\frac{i}{2}[\gamma^{\mu},\gamma^{\nu}] ,
\ea
\eeq
where $a^{\mu}$ and $\varepsilon_{\mu\nu}=-\varepsilon_{\nu\mu}$ are
the parameters of the transformation. In order to make the symmetry
local, it is sufficient to  replace the differentials by covariant
differentials (thus introducing the connection $\omega$), to replace the basis of one--forms $dx^{\mu}$ of $M$ by the cotetrad
basis $e^a$ on the resulting RC space $\mathcal{M}(e,\omega)$, and to use the Hodge star operator $\star$ adapted to
$\mathcal{M}$. The resulting Lagrangian four--form is
\be\label{LF0gr}
\ba
&\tilde{\mathfrak{L}}_{F0}=
-i\l{\star e_a}\r \wedge \ov{\psi}\gamma^aD\psi-m\ov{\psi}\psi\,\epsilon , \\
&D\psi=d\psi-\frac{i}{4}\omega_{ab}\Sigma^{ab}\psi 
\ea
\ee
(the matrices $\gamma^a$, $a=0,\dots,3$ are just the same as
$\gamma^{\mu}$, $\mu=0,\dots,3$). Here $\epsilon=e^0\wedge e^1\wedge
e^2\wedge e^3$ is the canonical volume element on $\mathcal{M}$. The
coupling procedure of this kind will be referred to as the minimal coupling procedure (MCP) for the
gravitational interaction. The one--forms $\omega_{ab}=-\omega_{ba}$,
which endow the space--time with the metric--compatible connection,
may be interpreted as gauge--fields corresponding to Lorentz
rotations. Although the relation of $e^a$ to the translational gauge--fields is more subtle, the procedure can be given interpretation in the framework of gauge theory of the Poincar{\'e} group (see \cite{Kazm3} for an exhaustive and simple treatment). In EC theory, the gauge--field--part of the Lagrangian is
 taken to be
 $\mathfrak{L}_G=-\frac{1}{4k}\epsilon_{abcd}e^a\wedge e^b\wedge
 \Omega^{cd}$, where $k$ is a constant and
 ${\Omega^a}_b=d{\omega^a}_b+{\omega^a}_c\wedge {\omega^c}_b$ the
 curvature two--form on $\mathcal{M}$.
It is crucial that the first--order formulation of general relativity
is much more adequate for gauge formulation than the standard
second--order one.  We shall now address the problem which the first--order approach entails.

Let (\ref{S}) denote the action functional of a classical field theory
in Minkowski space $M$.
It is well known that the transformation
\be\label{Lch}
\mathcal{L}\rightarrow\mathcal{L}'=\mathcal{L}+\partial_{\mu}V^{\mu}
\ee
of the Lagrangian density changes $\mathfrak{L}$ by a differential.
When introducing a new interaction, it seems reasonable to require
that the resulting theory be independent on whether we have added a divergence
to the initial Lagrangian density or not. Let us now specialize again to the Dirac field and consider the effect of the transformation (\ref{Lch}) of the initial Lagrangian on the final Lagrangian
four--form on $\mathcal{M}$. We shall consider the vector
field of the form 
\be\label{V}
V^{\mu}=aJ_{(V)}^{\mu}+bJ_{(A)}^{\mu}, \quad a,b\in\mathbb{C},
\ee
where $J_{(V)}^{\mu}=\ov{\psi}\gamma^{\mu}\psi$ and $J_{(A)}^{\mu}=\ov{\psi}\gamma^{\mu}\gamma^5\psi$ are the Dirac vector and
axial currents (this is the only possible form which is quadratic in $\psi$ and transforms as a vector under proper Lorentz transformations). It is straightforward to check that the following Leibniz rule applies
\be\label{Leib}
\l{D\ov{\psi}}\r
C^a\psi+\ov{\psi}C^aD\psi=d\l{\ov{\psi}C^a\psi}\r+{\omega^a}_b \l{\ov{\psi}C^b\psi}\r,
\ee
where $C^a:=a\gamma^a+b\gamma^a\gamma^5$. Hence under the minimal
coupling $d\psi\rightarrow D\psi$ the differential $dV^{\mu}$ of
(\ref{V}) will pass into $DV^a=dV^a+{\omega^a}_bV^b$. Using the
identity $\partial_{\mu}V^{\mu}d^4x=-\star (dx_{\mu})\wedge dV^{\mu}$
one can then conclude that the change in the resulting Lagrangian four--form on
$\mathcal{M}$ (under the transformation (\ref{Lch}) of the initial
Lagrangian density) will be
\be\label{TVtetr}
\mathfrak{L}'-\mathfrak{L}=d\l{V\lrcorner\epsilon}\r-T_aV^a\epsilon ,
\ee
where $T^a={T^{ba}}_b$ is the torsion trace (the components of
the torsion tensor in the tetrad basis are given by the equation
$\frac{1}{2}{T^a}_{bc}e^b\wedge e^c=de^a+{\omega^a}_b\wedge
e^b$) and $\lrcorner$ denotes the internal product. When deriving (\ref{TVtetr}), it
is necessary to use metricity of $\omega$. Within the framework of classical general relativity,
where the torsion of the connection is assumed to vanish, the
result would be again a differential. In EC theory the
torsion is determined by the spin of matter and does not vanish in general. Hence, the equivalent
theories of the Dirac field in flat space can lead to the
non--equivalent theories with gravitation. Surprisingly, this fact
has been used by many authors to remove a serious pathology of the Lagrangian (\ref{LF0gr}). This
Lagrangian is neither real, nor does it differ by divergence from the real
one. As a result, the equations obtained by varying with respect to
$\psi$ and $\ov{\psi}$ are not equivalent and together impose too
severe restrictions on the field. The commonly accepted solution is to adopt 
\be\label{LFR}
\mathfrak{L}_{FR}=
-\dfrac{i}{2}\l{\star dx_{\mu}}\r \wedge \l{ \ov{\psi}\gamma^{\mu}
  d\psi-\ov{d\psi}\gamma^{\mu}\psi}\r-m\ov{\psi}\psi d^4x
\ee
as an appropriate flat space Lagrangian ((\ref{LFR}) differs from
(\ref{LF0}) by a differential). The application of MCP yields
\beq
\tilde{\mathfrak{L}}_{FR}=
-\dfrac{i}{2}\l{\star e_a}\r \wedge \l{ \ov{\psi}\gamma^a
  D\psi-\ov{D\psi}\gamma^a\psi}\r-m\ov{\psi}\psi\,\epsilon . 
\eeq
This choice of Lagrangian served as the
basis for physical investigations in numerous papers. But the reality
requirement does not fix the theory uniquely. We can next add to
$\mathcal{L}_{FR}$ the
divergence of a vector field of the form (\ref{V}), where now the parameters
$a$, $b$ are required to be real, since we do not want to destroy
the reality of the Lagrangian. This may lead to the meaningful physical effects \cite{Kazm1,Kazm2}.
Hence, the standard MCP for first--order gravity appears to involve an ambiguity.
\section{How to remove the ambiguity?}\label{ambr}
For the Dirac field, the linear space of the representation of the
gravitational gauge group is $\mathbb{C}^4$ and the space $Ran(\pi)$ is
spanned by the matrices $\Sigma^{ab}$. The natural Lorentz invariant
scalar product
$\langle\phi,\psi\rangle_{\rho}=\phi^{\dag}\gamma^0\psi$ on
$\mathbb{C}^4$ induces the
product $\langle\langle
X,Y\rangle\rangle_{\rho}=trace\l{\gamma^0X^{\dag}\gamma^0Y}\r$ on
$Lin(\mathcal{V})$. For any representation of the matrices $\gamma^a$
that is unitarily equivalent to the Dirac representation, the orthogonal complement is
spanned by ${\bf 1}$, $\gamma^5$, $\gamma^a$, $\gamma^5\gamma^a$. Hence we have
\be
\ba
&\mathcal{D}\psi=D\psi+\mathbb{B}\psi,\\
&D\psi=d\psi+\mathbb{A}\psi, \quad
\mathbb{A}=-\frac{i}{4}\omega_{ab}\Sigma^{ab},\\
&\mathbb{B}=\chi{\bf 1}+\kappa\gamma^5+\tau_a\gamma^a+\rho_a\gamma^5\gamma^a,
\ea
\ee
where $\chi$, $\kappa$, $\tau_a$, $\rho_a$ are complex
valued one--forms on space--time. We will require that the Leibniz
rule hold for the Dirac vector and axial currents,
\beq
\ba
&(\mathcal{D}\ov{\psi})\gamma^a\psi+\ov{\psi}\gamma^a\mathcal{D}\psi=
dJ_{(V)}^a+\tilde{\omega}{^a}_bJ_{(V)}^b,\\
&(\mathcal{D}\ov{\psi})\gamma^a\gamma^5\psi+\ov{\psi}\gamma^a\gamma^5\mathcal{D}\psi=
dJ_{(A)}^a+\tilde{\omega}{^a}_bJ_{(A)}^b,
\ea
\eeq
where
$\mathcal{D}\ov{\psi}:=(\mathcal{D}\psi)^{\dag}\gamma^0$
and $\tilde{\omega}{^a}_b$ represents a modified
connection on the RC space. Straightforward
calculations show that these equations are satisfied if and only if
\beq
\tilde{\omega}{^a}_b={\omega^a}_b+\lambda\delta^a_b, \quad \mathbb{B}=\frac{1}{2}\lambda{\bf
  1}+i\mu_1{\bf 1}+i\mu_2\gamma^5,
\eeq
where $\lambda:=2Re\l{\chi}\r$, $\mu_1:=Im\l{\chi}\r$,
$\mu_2:=Im(\kappa)$ are real--valued one--forms.
Note that the one--forms $\mu_1$ and $\mu_2$ do not influence the resulting connection
on the RC space. If non--gravitational interactions were
included, the components of these
one--forms could be hidden in the gauge fields corresponding to the
localization of the global symmetry of the change of phase $\psi\rightarrow e^{i\alpha}\psi$ and the
approximate symmetry under the chiral transformation $\psi\rightarrow
e^{i\alpha\gamma^5}\psi$. In order not to involve non--gravitational
interactions, one needs to set $\mu_1$
and $\mu_2$ to zero.\par
According to the ideas presented at the beginning of this report,
$\lambda$ should be determined by $\omega$ and $e$ in such a way that it is a scalar
(compare (\ref{AandB}) and the remarks concerning the dependence of
$\mathbb{B}$ on $\mathbb{A}$ and $e$). What is more, the procedure is expected
to be free of the ambiguity. To see that all the requirements can be accomplished, note that the divergence 
$\partial_{\mu}V^{\mu}d^4x=-\star(dx_{\mu})\wedge dV^{\mu}$
will pass into $-\star(e_a)\wedge\l{dV^a+{\tilde{\omega}}{^a}_bV^b}\r$.
Hence, (\ref{TVtetr}) imply that the
procedure will yield unique results for generic $\omega$ if and only
if $\lambda=\mathbb{T}$, where $\mathbb{T}=T_ae^a$ is the
torsion--trace--one--form, which is indeed a scalar under local Lorentz
(or Poincar{\'e}) transformations $\omega\rightarrow
\Lambda\omega\Lambda^{-1}-d\Lambda\Lambda^{-1}$, $e\rightarrow\Lambda
e$. Hence, there exists precisely one coupling procedure which is
free of the ambiguity and satisfies all the requirements.

From the perspective of the base manifold $\mathcal{M}$, it seems that
the procedure could be stated briefly by saying that the modified
connection $\tilde{\omega}{^a}_b={\omega^a}_b+\mathbb{T}\delta^a_b$
should be used in MCP, instead of the original metric connection
$\omega$ entering $\mathfrak{L}_G$. However, it would not be clear
then how the new connection is to be implemented on spinors (the simple
substitution $\omega\rightarrow\tilde{\omega}$ in $D\psi$ would not
work well). What is more, there are other possibilities of modifying
the connection so that its application in MCP guarantees uniqueness. The
simplest way to achieve this would be to subtract the contortion tensor. This would
result in Levi--Civita connection reducing the formalism effectively
to the second--order one. The torsion would entirely disappear from
the theory. Less drastic possibility could be to retain only the
antisymmetric part of the torsion tensor by adopting $\tilde{\omega}_{ab}=\stackrel{\circ}{\omega}_{ab}-\frac{1}{2}T_{[abc]}e^c$,
where $\stackrel{\circ}{\omega}$ is the Levi--Civita part of $\omega$
and $T_{abc}$ the torsion of $\omega$. For the Dirac field, all such possibilities
necessarily violate one of the assumptions supporting our approach
(the two that were mentioned produce $\mathbb{B}$ that does not take
values in the orthogonal complement of $Ran(\pi)$ -- this makes
impossible reading out the connection $\omega$, that ought to be used in the
construction of $\mathfrak{L}_G$, from given
$\mathcal{A}=\mathbb{A}+\mathbb{B}(\mathbb{A},e)$). A different
approach is possible, in which the corrected connection takes values in an extension of the original
Lie algebra. One should specify what kind of
extensions are allowed, how the original connection is to be retrieved
from the extended one and to establish the dependence of Yang--Mills fields of the
extension from those of the original theory. In the case
discussed here, extending $so(1,3)$ by
dilatations would work well. However, the details of such an abstract
approach ought to be considered with care and this will not be done in
this brief report. 

The new connection $\tilde{\omega}$ on $\mathcal{M}$ is not
metric. One could hope that $\omega$ could be obtained from
$\tilde{\omega}$ as its metric part. This is however not the case. Let
us recall that the coefficients ${\Gamma^a}_{bc}$ of any connection can
be decomposed as
\be
\Gamma{^a}_{bc}=\stackrel{\circ}{\Gamma}{^a}_{bc}+K{^a}_{bc}+L{^a}_{bc},
\ee
where $\stackrel{\circ}{\Gamma}{^a}_{bc}$ is the Levi--Civita part
determined by the metric $g=\eta_{ab}e^a\otimes e^b$,
$K_{abc}:=\frac{1}{2}(T_{cab}+T_{bac}-T_{abc})$ the contortion and 
$L_{abc}=-\frac{1}{2}\l{\nabla_bg_{ca}+\nabla_cg_{ba}-\nabla_ag_{bc}}\r$
the nonmetricity. The contortion of $\tilde{\omega}$ is related to
that of $\omega$ by 
$\tilde{K}_{abc}=K_{abc}+\eta_{cb}T_a-\eta_{ca}T_b$. The metric
part of $\tilde{\Gamma}_{abc}$ is therefore equal to
$\Gamma_{abc}+\eta_{cb}T_a-\eta_{ca}T_b$, and not to
$\Gamma_{abc}$. 

\section*{Acknowledgements}
I wish to thank the referee of PRD for bringing the solution based on
anti--symmetric part of the torsion to my attention and for other
important remarks that improved this report. I am also grateful to Wojciech Kami{\'n}ski, Jerzy Lewandowski and Andrzej
Trautman for helpful comments. This work was partially supported by 
the 2007-2010 research project N202 n081 32/1844,
the Foundation for Polish Science grant ''Master''.

\end{document}